\documentclass[journal]{IEEEtran}
\usepackage{amsmath,amsfonts}
\usepackage{algorithmic}
\usepackage{array}
\usepackage[caption=false,font=normalsize,labelfont=sf,textfont=sf]{subfig}
\usepackage{textcomp}
\usepackage{stfloats}
\usepackage{url}
\usepackage{verbatim}
\usepackage{graphicx}
\usepackage{amsmath}
\usepackage{graphicx}
\usepackage{multirow}
\usepackage{mathtools}
\newcommand{\mlfix}{\mathtoolsset{multlined-width=0.9\displaywidth}}
\newcommand{\mlfull}{\mathtoolsset{multlined-width=\displaywidth}}
\usepackage{tikz}
\usepackage{soul}
\usetikzlibrary{arrows.meta}
\usetikzlibrary{angles,quotes}
\newcommand{\beq}{\begin{equation}}
\newcommand{\eeq}{\end{equation}}
\newcommand{\eref}[1]{(\ref{#1})}
\newcommand{\sub}[1]{_{\text{#1}}}

\newcommand{\micron}{\,\mu\mbox{m}}

\newcommand{\dd}{\delta}
\DeclareMathOperator{\csch}{csch}
\def\BibTeX{{\rm B\kern-.05em{\sc i\kern-.025em b}\kern-.08em
    T\kern-.1667em\lower.7ex\hbox{E}\kern-.125emX}}
\usepackage{balance}

\begin{document}
\title{Bounds on Distinguishing Separated Wires Using Magnetic Field Measurements}

\author{\IEEEauthorblockN{Adrian Mariano\IEEEauthorrefmark{1}, Jacob Lenz\IEEEauthorrefmark{1}, Dmitro Martynowych\IEEEauthorrefmark{1}\IEEEauthorrefmark{2}, Christopher Miller\IEEEauthorrefmark{1}, Sean Oliver \IEEEauthorrefmark{1}} \\
\IEEEauthorblockA{\IEEEauthorrefmark{1}Quantum Information Sciences, Optics, and Imaging Department, The MITRE Corporation, McLean, VA 22102 USA \IEEEauthorrefmark{2}Laboratory for Information and Decision Systems, Massachusetts Institute of Technology, Cambridge, MA 02139 USA}  \\ 
Corresponding Author: Adrian Mariano (email: adrian@mitre.org)}


\maketitle

\bibliographystyle{IEEEtran}

\begin{abstract}
Magnetic current imaging (MCI)
is useful for non-destructive characterization of microelectronics,
including both security analysis and failure analysis, because magnetic fields
penetrate the materials that comprise these components to enable
through-package imaging of chip activity.
Of particular interest are new capabilities offered by emerging
magnetic field imagers, such as the Quantum Diamond Microscope,
which provide simultaneous wide field-of-view, high spatial resolution
vector magnetic field imaging capabilities under ambient conditions.
While MCI offers several advantages for non-destructive measurement of
microelectronics functional activity, there are many limitations of the technique due to rapid falloff of magnetic fields and loss of
high frequency spatial information at large sensor standoff
distances. To understand spatial resolution as a function of standoff distance, 
we consider the
problem of using magnetic fields to distinguish (1) between a single wire 
carrying current $I$ and a pair of wires carrying current $I/2$ in the same 
direction and (2) between no currents and a pair of wires carrying current $I/2$ in opposite 
directions.  In both cases, we compare performance for a single point
measurement, representative of typical magnetometers, to
performance for an array of measurements found in emerging magnetic
imaging devices.   Additionally, we examine the advantage provided by measurement of the full
vector magnetic field compared to measurement of a single component.
We establish and compare for the first time the theoretical lower bounds
on separability based on the wire separation and sensor standoff 
distance of the magnetic field measurements obtained from traditional
and new microelectronics reliability tools.
\end{abstract}


\section{Introduction}
\IEEEPARstart{A}{ssuring} microelectronics security is becoming increasingly challenging as integrated circuits (ICs) utilize smaller feature sizes, heterogeneous integration, and three-dimensional (3D) die stacking.  Counterfeit ICs and hardware Trojans can be difficult to identify due to the potentially limited and area-specific activity they introduce. Non-destructively detecting evidence of tampering is often beneficial for the preservation of time and resources depending on the testing environment and availability of samples. The advances in IC technology introduce the need for characterization tools that can detect anomalous activity non-destructively and, in some cases, in 3D to ensure devices are not counterfeit and are free from hardware Trojans \cite{Guin2014, GuinU2014, Xue2020}. 

Microelectronics security verification relies on a suite of tools that each have their own advantages and drawbacks. Established tools for the detection of counterfeit ICs and hardware Trojans include optical imaging \cite{Ghosh2019}, thermal mapping \cite{Ghosh2020}, X-ray imaging \cite{Mahmood2015}, signature detection with resonance cavities \cite{Nechiyil2023}, and logic/output testing \cite{Chakraborty2009,Jha2008}. Optical imaging is used for rapid detection of abnormalities in IC package markings, which can be an indicator of counterfeit production \cite{Ghosh2019, Asadizanjani2017}. While this technique is low in cost and fast, it does not provide information about the component's inner structure and function and so will not expose a well-marked counterfeit chip. Thermal mapping provides information about the internal activity of ICs and can localize defects in 3D \cite{Frazier2016}. However, this method struggles with non-resistive failures like opens, whose detection requires the introduction of RF signals \cite{VanVeenhuizen2023}.  Heat dissipating structures and multi-material stack, which are common in 3D ICs and heterogeneous integrated chips, also complicate thermal mapping. X-ray imaging can accurately map an IC's internal circuitry to uncover differences from the expected die layout \cite{Mahmood2015}, but this instrumentation is expensive, does not provide information about chip function, and may not detect aged defects that can be a marker for counterfeits. The use of resonance cavities to measure electromagnetic signatures has demonstrated promise for counterfeit detection as well \cite{Nechiyil2023}. This sensor has excellent sensitivity to small changes in the measured chip but requires further development to ensure consistent results among different setups. Logic/output testing is a quick and cheap way to confirm that the chip is functional, potentially detecting malfunctioning counterfeits or revealing the presence of hardware Trojans. This method typically does not provide spatial information on the defect or alteration and will require other testing to determine the alteration's location and intended functionality \cite{Chakraborty2009,Jha2008}.

Magnetic current imaging (MCI) is a promising technique that addresses the limitations of current tools and adds to the suite of instruments used for IC characterization \cite{knauss,vallett2011,gaudestad}. To perform MCI, one first measures the magnetic fields induced by currents flowing through a functioning IC, and then one uses those measurements to analyze current paths. The distinct advantage of characterizing ICs through their magnetic field emissions is that these fields pass through the materials that make up ICs, and so they can be measured through device packaging and multiple layers of material. To date, scanning superconducting quantum interference devices (SQUIDs) have been the primary tool for MCI of ICs and have been shown to have sub-micron spatial resolution and exceptional noise characteristics \cite{Kirtley2016,Li2021a}. However, this tool has its own drawbacks. Measurement can be time-consuming as the SQUID magnetometer makes single point measurements that must be raster scanned to produce an image. Additionally, the instrument requires cryogenics, which poses logistical challenges and adds experimental complexity. Despite these drawbacks, the benefits offered by MCI have made it a compelling area of exploration in the ongoing evolution of IC characterization techniques. 

The Quantum Diamond Microscope (QDM) has recently emerged as a novel tool for MCI. This instrument uses magnetically-sensitive nitrogen-vacancy center quantum defects in a lab-grown diamond wafer to provide wide-field, vector magnetic field imaging of ICs under ambient conditions, eliminating the need for cryogenics \cite{Levine2019,Turner2020}. The QDM has achieved micrometer-scale spatial resolution of magnetic fields, and when the QDM is placed in contact with a chip, the current distributions have been recovered at that resolution \cite{Basso2022}.  Measurement resolution has been further improved through super-resolution techniques \cite{Mosavian2023}. The QDM allows for the correlation of spatial and temporal dynamics of a device under test \cite{Tang2023}, and initial demonstrations have shown that this can potentially be used for detection of Trojan hardware \cite{Ashok2022}. Additionally, in contrast to scanning SQUIDs that only measure one vector component of the magnetic field, the QDM can simultaneously measure all vector components of a magnetic field to enable a user to capture signals previously undetectable with a SQUID \cite{Oliver2021}. Given this unique combination of capabilities, the QDM represents a significant advancement in the field and offers the potential for more comprehensive and efficient IC characterization.

Because of the promise of emerging MCI instruments like the QDM, it is crucial to understand their spatial resolution limits as this dictates performance for non-destructive characterization of advanced ICs.  Previous studies have explored the limits of spatial resolution for single point measurement magnetic imaging tools such as the scanning SQUID magnetometer \cite{Wellstood}.  Here, we investigate the theoretical limits of spatial resolution for wide-field MCI techniques. We utilize first principles calculations to model the expected magnetic fields from simple current path geometries and establish the theoretical lower bounds for differentiating wires at various standoff distances. Furthermore, we assess the effects of noise on the ability to distinguish wires at various sensor standoff distances and current levels. This comprehensive theoretical analysis builds an understanding of wide-field MCI techniques and paves the way for their application in IC characterization.

\section{Problem Formulation}\label{sec:formulation}

When currents flow in wires, they create magnetic fields.  We would like to measure those magnetic fields and then solve the
inverse problem to estimate the current distribution.
It is important to understand the fundamental limits of magnetic field measurements
for determining those current distributions.  In this paper we provide a simple bound for the
simplest possible discrimination problem in the case
of infinite wires.  

To accomplish this, we consider the  simplest possible MCI task---distinguishing
two infinite current carrying wires from one wire or no wire---and derive conditions
under which this is possible. If this task is impossible, then it is a reasonable assumption that MCI cannot yield
useful results for more complicated and realistic problems under the
same conditions. We parameterize the wire sensing problem by:
$I$, the current on the wires, $z$, the vertical standoff between the
measurement device and the wire, $s$, the spacing between the two
wires, $\sigma$, the measurement noise, and the number and spacing of magnetic field measurements.

We assume that all wires have circular cross section and infinite extent
in the $y$ direction, and we assume that positive direction of current
through the wires is in the $y+$ direction.
When an infinite wire has a circular cross section we can exactly
compute its magnetic field using Ampere's law, independent of the
wire's radius, so our results are exact for any radius, but 
will be approximate in the case where wires deviate substantially from
circular cross section.  Because wires are infinite in the $y$
direction, the magnetic field in the $y$ direction is always zero.

In the \emph{parallel}
case we compare a single wire carrying current, $I$, to a pair of
wires each carrying current, $I/2$.  In the far field, these two
scenarios are indistinguishable.  What wire separation is needed for magnetic
field measurements to distinguish these two cases?

In the \emph{anti-parallel} case we compare the case of empty space to
a pair of wires, but in this case the currents run in opposite
direction, $I/2$ in one wire and $-I/2$ in the other.  Again, the
scenarios are indistinguishable in the far field, and we ask what wire
separation is necessary for magnetic field measurements to distinguish
these cases.

To assess the concept of detectability, consider the parallel case, where we must distinguish one wire carrying $I$ current, which produces magnetic field $B_1$, from two closely spaced wires carrying current $I/2$, which produce magnetic field, $B_2$.  If the magnetic field measurements are  corrupted by mean-zero Gaussian white noise ($\Sigma = \sigma^2 I$), then optimal discriminator is to compute the $L^2$ distances between the acquired signal $B_1$ and the acquired signal and $B_2$ \cite{duda}. The error rate of the detector is proportional to the Malhalanobis distance between $B_1$ and $B_2$: 
\beq 
   d_m = \sqrt{(B_1-B_2)^T\Sigma^{-1}(B_1-B_2)} = ||B_1 - B_2||/\sigma. 
\eeq
   Our approach is to derive bounds of this distance as a function of noise and the other measurement parameters.

\section{Parallel Case: Single Wire vs. Two Wires}\label{sec:parallel}

\begin{figure}
 \centerline{
  \begin{tikzpicture}[scale=1.3,>=Stealth]
    \draw[thin,<->,gray!80] (-3,0) -- (3,0) node [right] {$x$};
    \draw[thin,<->,gray!80] (0,-1) -- (0,2); 
    \draw[<->] (-1.5,.03) -- node [left]{$z$} (-1.5,1.47);
    \draw[<->] (-2.47,1.5) -- node[above] {$\Delta x$} (-2.03,1.5);
    \foreach \x in {-2.5,-2,...,2.5} \draw[fill=red] (\x,1.5) circle(.02);
    \draw[fill=yellow] (30:-1.2) circle(.07) -- (30:1.2) circle(.07);
    \draw (15:.54) node {$\theta$};
    \draw[<->] (120:.2) -- node[above,sloped] {$s$} +(30:1.2);
    \draw[blue] (30:1.2) -- node[below]{$x-s\cos\theta$}(2.5,1.5 |- 30:1.2) --
    node[left]{$z-s\sin\theta$} (2.5,1.5) node[anchor=south] {$(x,z)$};
  \end{tikzpicture}}
\caption{\label{geom} Two wires, shown in yellow, are each distance
  $s$ from the origin at angle $\theta$.  Measurements are taken at
  the red points on the line distance $z$ above the origin, spaced
  $\Delta x$ apart.  In blue are the distances needed to compute the
  $B$ field at the rightmost measurement point, which we assume is at
  $(x,z)$.  
}
\end{figure}
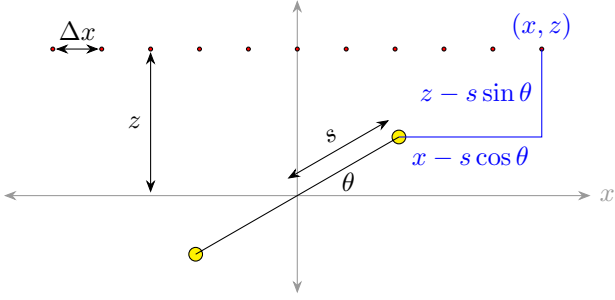

Assume that an infinite wire with circular cross section runs along
the $y$ axis carrying current, $I$, where as noted above, we always
assume positive direction of current flow is in the $y+$ direction.
We make $2N+1$
vector magnetic field measurements at the points, $(i\Delta x, z)$ for $i\in
\{-N,\ldots,N\}$.
The magnetic field at $(x,0,z)$ is given by
\beq  \label{onewire}
\begin{aligned}
       B_x&= \frac{\mu_0 I}{2\pi} \frac{z}{x^2+z^2}\\
       B_z&= \frac{-\mu_0 I}{2\pi} \frac{x}{x^2+z^2}.\\      
\end{aligned}
\eeq
Now consider the contrasting situation where two wires are present,
symmetrically located at distance, $s$, from the origin, rotated by
angle, $\theta$, as shown in 
Figure~\ref{geom}.
The field due to the pair of wires, each carrying current $I/2$, is given
by 
\beq
\begin{aligned}   \label{fullangle}  \mathtoolsset{multlined-width=0.9\displaywidth}
   \begin{multlined}B_x=\frac{\mu_0 I}{4 \pi} \left( \frac{z-s\sin\theta}{(x-s\cos\theta)^2+(z-s\sin\theta)^2}\right.\\ 
           {} + \left.\frac{z+s\sin\theta}{(x+s\cos\theta)^2+(z+s\sin\theta)^2} \right),\end{multlined}\\
           \mathtoolsset{multlined-width=0.9\displaywidth}
   \begin{multlined}B_z = \frac{-\mu_0 I}{4 \pi} \left( \frac{x-s\cos\theta}{(x-s\cos\theta)^2+(z-s\sin\theta)^2}\right.\\ +\left. \frac{x+s\cos\theta}{(x+s\cos\theta)^2+(z+s\sin\theta)^2} \right).\end{multlined}
\end{aligned}
\eeq
We want to know the smallest possible value of $s$ where we can still
distinguish the wires from each other, so it is appropriate to use a
small $s$ approximation.  
If we multiply the numerator and denominator in~\eqref{fullangle} by $1/z^2$
then in the resulting expression, $s$ always appears as $s/z$.  Therefore we can use an expansion in $s$ which
will be valid for all $x$ and for $s\ll z$.  This expansion gives
\beq
\begin{aligned}  \label{approxangle} \mathtoolsset{multlined-width=\displaywidth}
  \begin{multlined}B_x \approx \frac{\mu_o I}{4 \pi}
  \left(
      \frac{2 z}{x^2+z^2}  \right.\\\left.+\frac{2s^2\left(z(3x^2-z^2) \cos2\theta -x(x^2-3z^2)\sin2\theta\right)}
           {(x^2+z^2)^3}\right),\end{multlined}\\ \mathtoolsset{multlined-width=\displaywidth}
           \begin{multlined} 
 B_z \approx\frac{-\mu_o I}{4 \pi}
 \left(
 \frac{2 x}{x^2+z^2} \right.\\+\left.\frac{2s^2\left(x(x^2-3z^2)\cos2\theta - z(z^2-3x^2)\sin2\theta\right)}
           {(x^2+z^2)^3}\right). \end{multlined}
\end{aligned}
\eeq
Note that the first term in the expansion matches the field for the
single wire case given in \eqref{onewire}.
Therefore, the difference between the one and two wire cases is given
by the second term alone.  Rewriting $\Delta x$ relative to $z$
simplifies the equations, so let $\delta=\Delta x/z$.  In the second
term we then make the substitution $x=i\Delta x=i\delta z$ to obtain
\beq
\begin{aligned}  \mlfix
  \begin{multlined}
  \Delta B_x \approx \left(\frac{\mu_0 I}{2\pi}\right)\left( \frac{s^2}{z^3}\right)\\\cdot\left(
    \frac{(3i^2\dd^2-1)\cos2\theta - i\dd(i^2\dd^2-3)\sin2\theta}{(i^2\dd^2+1)^3}\right),\end{multlined}\\
    \mlfix \begin{multlined}
    \Delta B_z \approx \left(\frac{-\mu_0 I}{2\pi}\right)\left( \frac{s^2}{z^3}\right)\\\cdot\left(
    \frac{i\dd(i^2\dd^2-3)\cos2\theta + (3i^2\dd^2-1)\sin2\theta}{(i^2\dd^2+1)^3}\right).\end{multlined}
\end{aligned}
\eeq

The squared $L^2$ distance between the one wire and two wire measurements over the
full vector valued magnetic field measurement is given by 
\beq
   d\sub{full}^2 = \sum_{i=-N}^{N} \big(\Delta B_x(i)\big)^2 + \sum_{i=-N}^{N} \big(\Delta B_z(i)\big)^2.
\eeq
Substituting in the approximation from~\eqref{approxangle} gives
\beq
\begin{multlined} \mlfull
  d\sub{full}^2 \approx \left(\frac{\pm\mu_0 I}{2\pi}\right)^2\left( \frac{s^2}{z^3}\right)^2 \\ \cdot \left[\sum_{i=-N}^{N}\left(
  \frac{(3i^2\dd^2-1)\cos2\theta - i\dd(i^2\dd^2-3)\sin2\theta}{(i^2\dd^2+1)^3}\right)^2\right.\qquad
  \\
  +\left.\sum_{i=-N}^{N}
  \left(\frac{i\dd(i^2\dd^2-3)\cos2\theta + (3i^2\dd^2-1)\sin2\theta}{(i^2\dd^2+1)^3}\right)^2\right].\\
\end{multlined}
\eeq
Note that the coefficients on $\sin2\theta$ and $\cos2\theta$ match
across the two terms.  The result is that the cross terms in
$\cos2\theta\sin2\theta$ will cancel, and the squared terms will
combine and simplify leading eventually to
\beq  \label{full_sum}
  d\sub{full}^2 \approx \left(\frac{\mu_0 I}{2\pi}\right)^2\left( \frac{s^2}{z^3}\right)^2 \sum_{i=-N}^{N}
  \frac{1}{(i^2\dd^2+1)^3}.
  \eeq
  It is surprising that this result is independent of $\theta$ since the problem lacks rotational symmetry.
  This happens because the dependence on 
  $\theta$ resides entirely in the higher order terms.

If we are only able to measure the magnetic field in the $z$ direction
then the distance expression remains dependent on $\theta$:  
\begin{align} 
  d_z^2 &= \sum_{i=-N}^{N} \big(\Delta B_z(i)\big)^2\\  \label{zonly_sum}
      &\mlfix \begin{multlined}\approx \left(\frac{-\mu_0 I}{2\pi}\right)^2\left( \frac{s^2}{z^3}\right)^2\\
  \cdot\sum_{i=-N}^{N}
  \left(\frac{i\dd(i^2\dd^2-3)\cos2\theta + (3i^2\dd^2-1)\sin2\theta}{(i^2\dd^2+1)^3}\right)^2. \end{multlined}
\end{align}

Let $\eta$ be the square root of the summation term in the squared distance expression.  For the full vector-valued
$B$ field measurement, $\eta$ is defined using the summation
from~\eqref{full_sum} giving
\beq  \label{eta_expression}
    \eta =\eta\sub{full}= \sqrt{\sum_{i=-N}^N \frac{1}{(i^2\dd^2+1)^3}},
\eeq
and for the $B_z$ measurement the definition is from~\eqref{zonly_sum}, giving
\begin{align}
   \eta\ &= \eta_z \\&= \sqrt{
  \sum_{i=-N}^{N}
  \left(\frac{i\dd(i^2\dd^2-3)\cos2\theta + (3i^2\dd^2-1)\sin2\theta}{(i^2\dd^2+1)^3}\right)^2}.
\end{align}
We then have for the distance,
\beq
   d = \frac{\mu_0 I}{2\pi} \frac{s^2\eta}{z^3}.
\eeq
In order to declare the wires detectable we require $d>\alpha\sigma$,
where $\alpha$ is a parameter that determines the false alarm rate of
the detector. This leads to the requirement that
\beq   \label{multibound}
s>\sqrt{\frac{2 \alpha\sigma\pi z^3}{\mu_0I\eta}}
\qquad\text{or}\qquad
z<\sqrt[3]{\frac{\mu_o I \eta s^2}{2\pi\alpha\sigma}}.
\eeq
In the case of a single centered measurement, $\eta\sub{full}=1$, giving a simple bound
for that single measurement case.  The result for $\eta_z$ depends on angle, with $\eta_z=|\sin 2\theta|$.  
This result indicates that detection is best around $\theta=45^\circ$ but degrades for other angles and is impossible when $\theta=0^\circ$ or
$\theta=90^{\circ}$.  This makes sense because with a single, centered
measurement, the $\Delta B_z$ value is zero when $\theta\in\{0^\circ,90^\circ\}$; without $\Delta B_x$ it is
impossible to distinguish one wire from two without an off-center measurement.  

To provide a useful bound with
multiple measurements, we need to estimate $\eta$ for $N>0$. If we allow $N\rightarrow\infty$ then the
infinite sums can be solved exactly.  Adding measurements at
infinity will not significantly affect the result because the
magnetic field differences between a single wire and two wires is 
tiny at locations far away from the wires.  

Using Mathematica we find that
\beq \mlfix
\begin{multlined}  \left(\eta\sub{full}^\infty\right)^2=  \label{hyper} 
\sum_{i=-\infty}^{\infty} \frac{1}{(i^2\dd^2+1)^3} =
  \frac{3\pi}{8\dd}  \coth(\pi/\dd) \\ \qquad\qquad\qquad\qquad +\frac{3 \pi^2}{8\dd^2} \csch^2(\pi/\dd)\\
  + \frac{\pi^3}{4\dd^3} \coth(\pi/\dd)\csch^2(\pi/\dd).
\end{multlined}\eeq
For small $\dd$ we observe that 
$\coth(\pi/\dd)\approx1$ and additionally $\csch(\pi/\dd)\approx0$.
Making these substitutions in~\eqref{hyper} leads to the approximation
\beq  \label{etafullfinal}
\eta\sub{full}^{\infty} \approx \sqrt{\frac{3 \pi}{8\dd}}.
\eeq
When $\dd=1$ the relative error of this approximation for $\eta$ is
$-3.6\%$.
\begin{figure}
  \centerline{
    \includegraphics[width=3.5in]{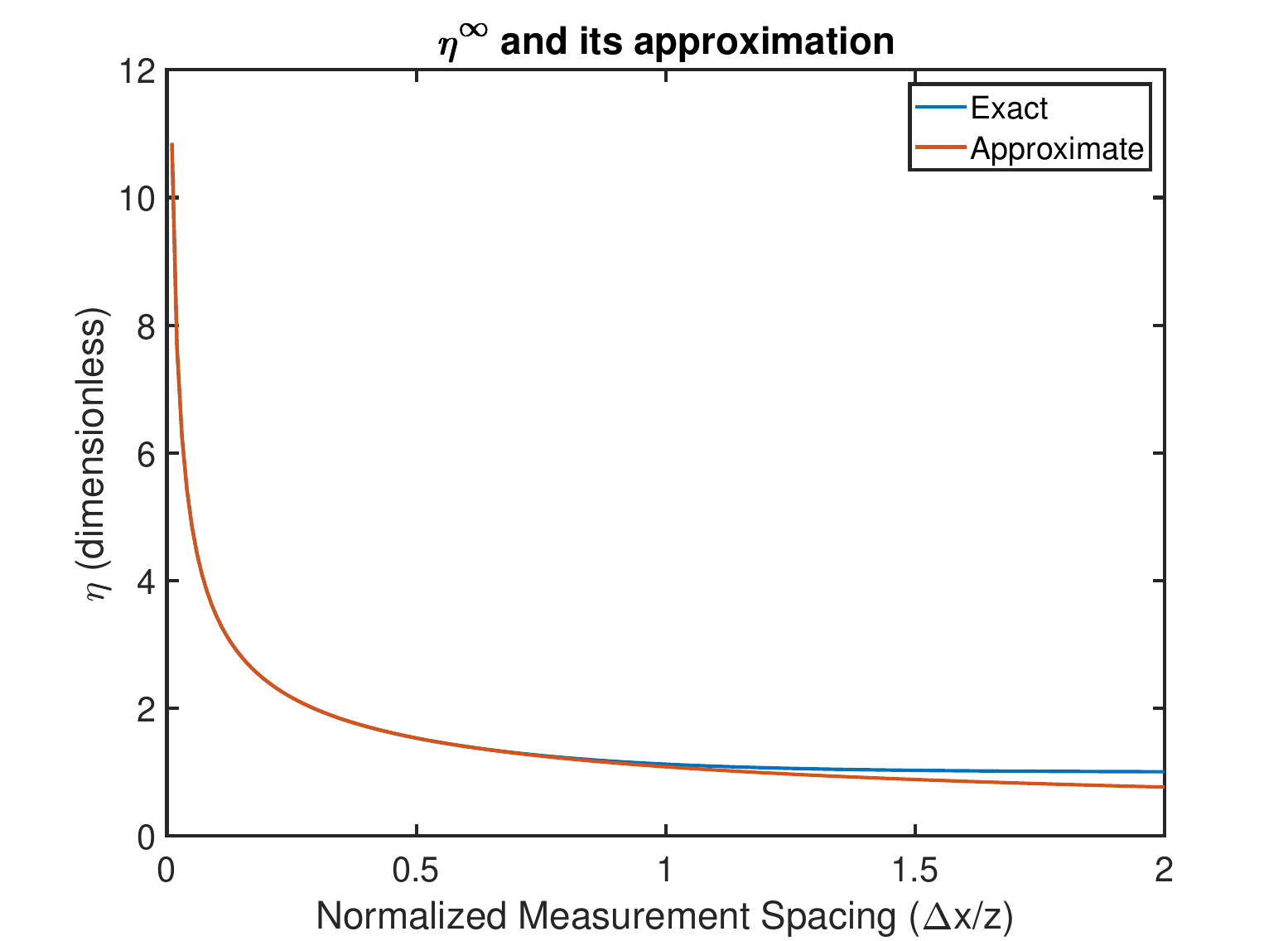}}
    \vspace*{4mm}
   \centerline{ \includegraphics[width=3.5in]{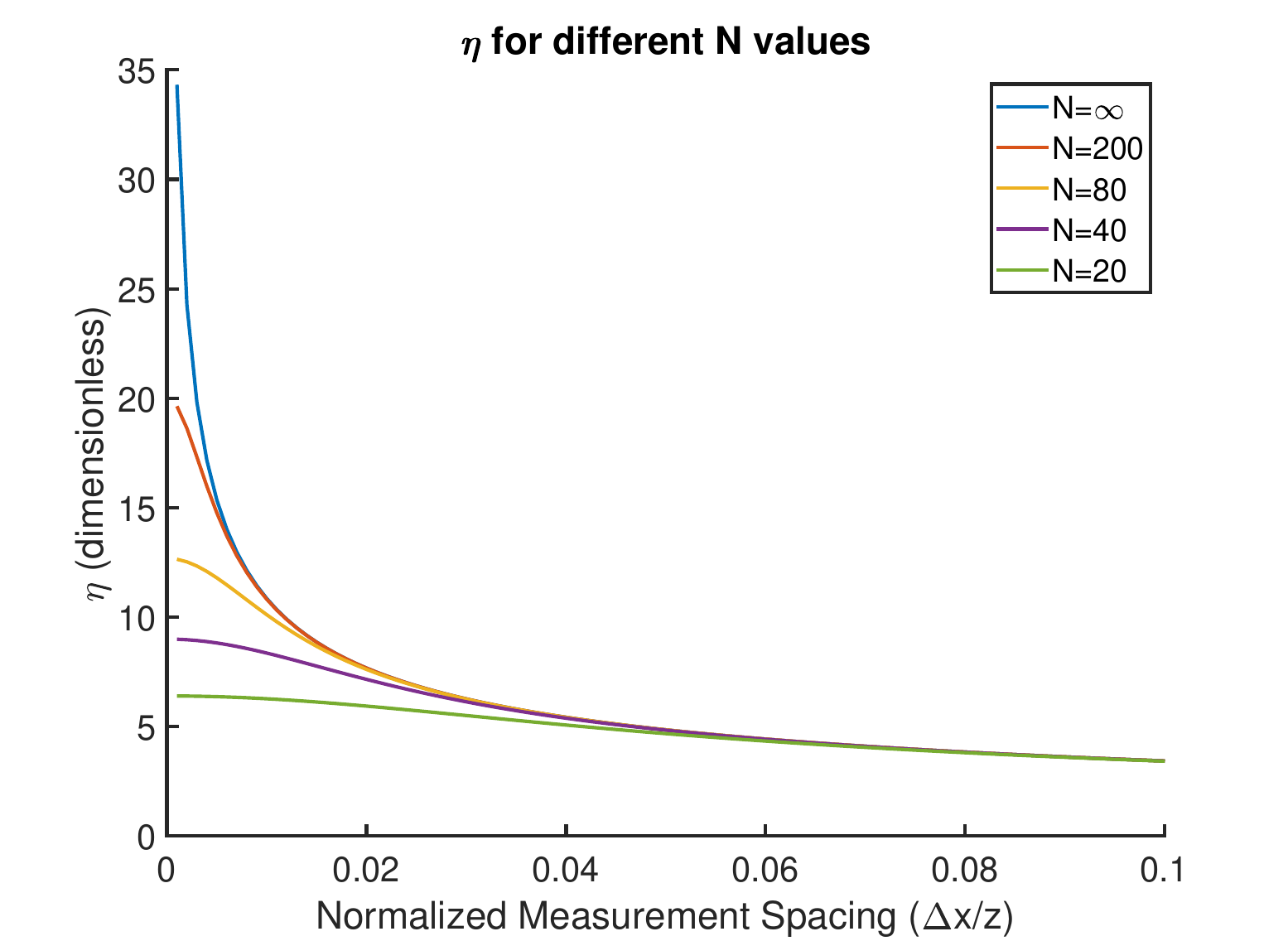}}
\caption{\label{etafig}
  The top plot shows the value of $\eta\sub{full}^{\infty}$
  compared to its approximation from~\eqref{etafullfinal} as a
  function of normalized measurement spacing, $\dd$.  The bottom
  plot shows $\eta\sub{full}$ for different values of $N$ compared to
  the exact limiting case for $N\rightarrow\infty$.  Note that the
  normalized measurement spacing, $\dd$, must be quite small before
  the curves diverge from each other, so the $N\rightarrow\infty$
  approximation has broad applicability. 
}
\end{figure}
Figure~\ref{etafig} shows the value of $\eta\sub{full}^{\infty}$
compared to its approximation, and compared to values with finite
$N$.  

When $\dd$ is large, $\coth(\pi/\dd)\approx \dd/\pi$ and
$\csch(\pi/\dd)\approx \dd/\pi$.  The terms in~\eqref{hyper} collapse giving the
result $\eta=1$.
This is reasonable because when $\dd$ is
sufficiently large only the central measurement matters---the other measurements are too far away from the
wires to contribute significant information.

We now consider the more complicated situation arising if we are only
able to make a scalar measurement of the $z$ component of the magnetic field.  In this case
Mathematica reports that $\left(\eta_z^\infty\right)^2$ is given by 
\beq
\begin{aligned}
  \sum_{i=-\infty}^{\infty}&
  \left(\frac{i\dd(i^2\dd^2-3)\cos2\theta +
    (3i^2\dd^2-1)\sin2\theta}{(i^2\dd^2+1)^3}\right)^2 =  \\
  &\frac{3 \pi}{16\dd} \coth(\pi/\dd) +\frac{3 \pi^2}{3\dd^2}  \csch^2(\pi/\dd)\\
  &\qquad\qquad+ \frac{\pi^3}{8\dd^3} \coth(\pi/\dd)\csch^2(\pi/\dd)\\
  &-\frac{\pi^6 \cos4\theta}{240\dd^6} \Big(
         16\coth^4(\pi/\dd)\csch^2(\pi/\dd)\\
   &  \qquad\qquad\qquad    +88\coth^2(\pi/\dd)\csch^4(\pi/\dd)\\
&  \qquad\qquad\qquad       +16\csch^6(\pi/\dd)\Big).
\end{aligned}\eeq
Using the same small $\dd$ approximation scheme, we set the terms containing
$\csch(\pi/\dd)$ to zero and set $\coth(\pi/\dd)$ to one,
which gives the result
\beq
\eta_z^{\infty} \approx \sqrt{\frac{3\pi}{16\dd}},
\eeq
exactly half the value found in~\eqref{etafullfinal} for
$\eta\sub{full}^{\infty}$.  The approximation is not as good, however.
When $\dd=1$ the relative error is $+116\%$.  At $\dd=0.5$ the relative
error is $+2.5\%$. For large $\delta$ we find that the terms collapse giving
\beq
\lim_{\delta\rightarrow\infty}\eta_z^{\infty}=\sqrt{\frac{1-\cos 4\theta}{2}} = |\sin 2\theta|,
\eeq
which agrees with our previous finding for the single measurement case.

\section{Anti-Parallel Case: No Wire vs. Two Wires}\label{sec:antiparallel}

For the anti-parallel case, we need to distinguish no current from the
current distribution due to two parallel wires in the $y$ direction each carrying current $I/2$ in opposite
directions.  In this case the expression for $B$ is
\beq
\begin{aligned}   \mlfull
   \begin{multlined}B_x = \frac{\mu_0 I}{4 \pi} \left( \frac{z-s\sin\theta}{(x-s\cos\theta)^2+(z-s\sin\theta)^2}\right.\\\left.
            - \frac{z+s\sin\theta}{(x+s\cos\theta)^2+(z+s\sin\theta)^2} \right),\end{multlined}\\
   \mlfull\begin{multlined}B_z = \frac{-\mu_0 I}{4 \pi} \left( \frac{x-s\cos\theta}{(x-s\cos\theta)^2+(z-s\sin\theta)^2}\right.\\\left. - \frac{x+s\cos\theta}{(x+s\cos\theta)^2+(z+s\sin\theta)^2} \right).\end{multlined}\\
\end{aligned}
\eeq
As before we use an expansion valid for $s\ll z$ to obtain
\beq
\begin{aligned}  \label{nowapprox}
   B_x &\approx \left(\frac{\mu_0 I}{2 \pi}\right) \frac{2sx z \cos\theta +s(z^2-x^2)\sin\theta}{(x^2+z^2)^2},  \\
   B_z &\approx \left(\frac{-\mu_0 I}{2 \pi}\right) \frac{ x(z^2-x^2) \cos\theta + 2sxz\sin\theta}{(x^2+z^2)^2}.
\end{aligned}
\eeq
When no wire is present, the $B$ field is zero,
so~\eqref{nowapprox} directly gives $\Delta B$ for this scenario. 
We again let $\Delta x = \delta/z$ and make the substitution $x=i\dd z$ to
obtain
\beq
\begin{aligned}
  \Delta B_x & \approx \left(\frac{\mu_0 I}{2 \pi}\right) \left(\frac{s}{z^2}\right)
              \frac{2i\dd\cos\theta +(1-i^2\dd^2)\sin\theta}{(i^2\dd^2+1)^2},\\   
  \Delta B_z &\approx \left(\frac{-\mu_0 I}{2 \pi}\right)  \left(\frac{s}{z^2}\right)
              \frac{(1-i^2\dd^2)\cos\theta -
                2i\dd\sin\theta}{(i^2\dd^2+1)^2}.\\
\end{aligned}              
\eeq
The distance expression for the full magnetic field undergoes a
simplification similar to the one that occurred in the parallel case
to give
\begin{align} 
   d\sub{full}^2 &= \sum_{i=-N}^{N} \big(\Delta B_x(i)\big)^2 + \sum_{i=-N}^{N} \big(\Delta B_z(i)\big)^2\\
   &= \left(\frac{\mu_0 I}{2\pi}\right)^2 \left(\frac{s}{z^2}\right)^2
   \sum_{i=-N}^N\frac{1}{(i^2\dd^2+1)^2}, \label{zetasumfull}
\end{align}
and also as before, no simplification is possible when only $B_z$ is
available:
\begin{align}
   d_z^2 &= \sum_{i=-N}^{i=N} \big(\Delta B_z(i)\big)^2\\ \label{zetasumz}
   &= \left(\frac{\mu_0 I}{2\pi}\right)^2 \left(\frac{s}{z^2}\right)^2
   \sum_{i=-N}^N \left(
     \frac{(1-i^2\dd^2)\cos\theta -
                  2i\dd\sin\theta}{(i^2\dd^2+1)^2}\right)^2.
\end{align}
Let $\zeta$ be the square root of the summation term from
either~\eqref{zetasumfull} or~\eqref{zetasumz}, so for the full
magnetic field we have from~\eqref{zetasumfull}
\beq  \label{zeta_expression}
\zeta = \zeta\sub{full} = \sqrt{\sum_{i=-N}^N \frac{1}{(i^2\dd^2+1)^2}},
\eeq
and for the $z$ only measurement from~\eqref{zetasumz} we have
\beq  \label{zeta_def}
\zeta = \zeta_z = \sqrt{
   \sum_{i=-N}^N \left(
     \frac{(1-i^2\dd^2)\cos\theta -
                  2i\dd\sin\theta}{(i^2\dd^2+1)^2}\right)^2}.
\eeq
The distance is given by
\beq
d = \frac{\mu_0 I}{2\pi} \frac{s\zeta}{z^2}
\eeq
and the requirement that $d>\alpha\sigma$ leads to
\beq  \label{sbound2}
s>\frac{2 \pi \alpha \sigma z^2}{\mu_0 I \zeta}
\qquad\text{or}\qquad
z<\sqrt{\frac{\mu_0I\zeta s}{2\pi\alpha\sigma}}.
\eeq

As before we can consider the case of a single measurement at directly over the wire at $x=0$, where we find that $\zeta\sub{full}=1$ and $\zeta_z=|\cos\theta|$, which gives a simple bound applicable to those cases.  When we only measure $B_z$, we get the best detection when the wires are separated
horizontally $\theta=0^\circ$) and we fail to detect vertically separated
wires ($\theta=90^\circ$) because $\Delta B_z$ is zero for the vertically separated case.  

Computing the sum for the infinite case with the full $B$ field reveals
\beq
\begin{aligned}
 \left(\zeta\sub{full}^\infty\right)^2=
\sum_{i=-\infty}^{\infty} \frac{1}{(i^2\dd^2+1)^2} &= \frac{ \pi \dd \coth(\pi/\dd) + \pi^2\csch^2(\pi/\dd)}{2\dd^2}
\\&\approx \frac{\pi}{2\dd}, 
\end{aligned}
\eeq
where we make the same small $\dd$ approximation as before.  This gives
\beq \label{zetafullfinal}
 \zeta\sub{full}^{\infty} \approx \sqrt{\frac{\pi}{2\dd}},
\eeq
which has an error of $-1.3\%$ when $\dd=1$. 
\begin{figure}
  \centerline{  \includegraphics[width=3.5in]{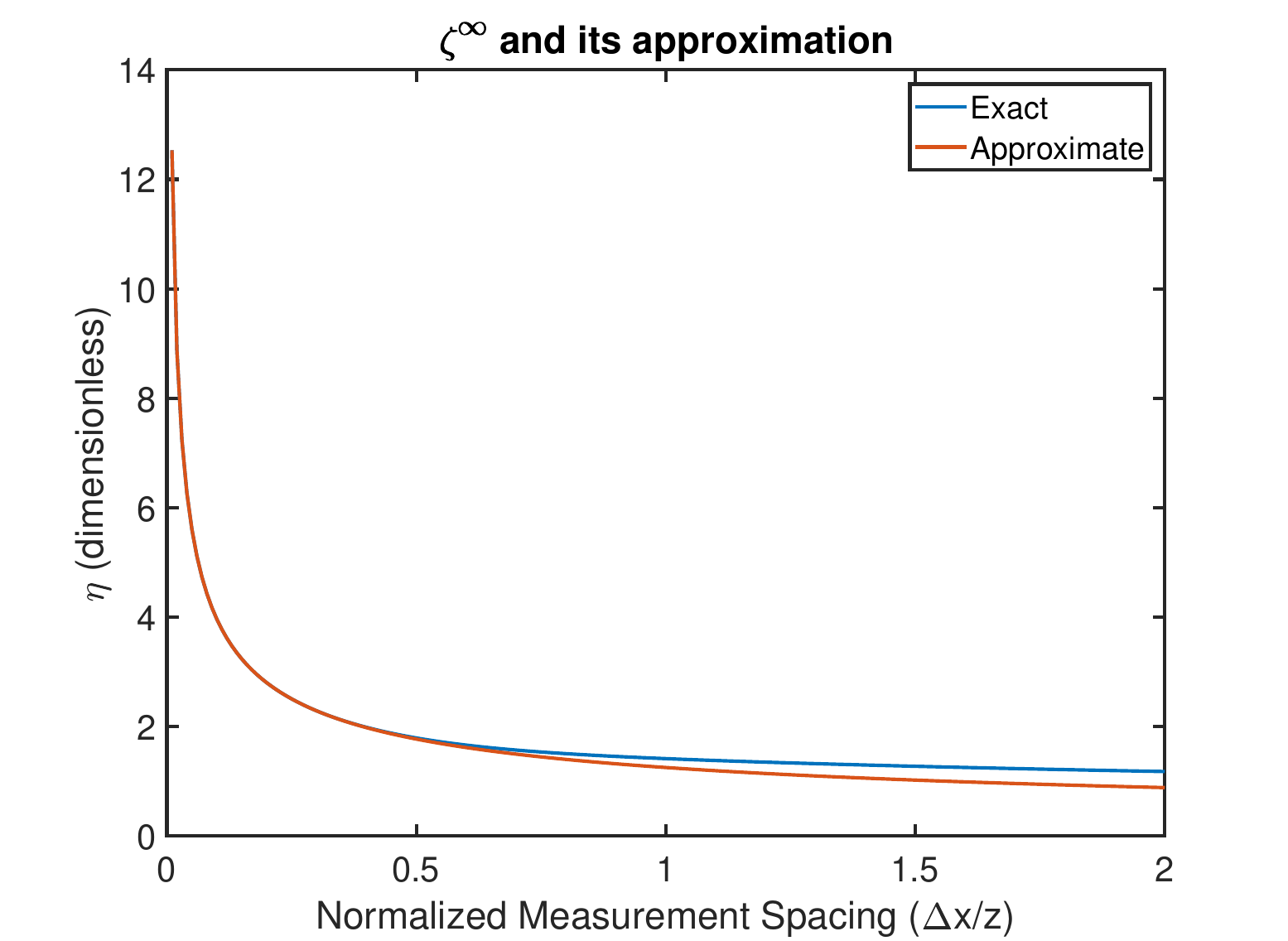}}
    \vspace*{4mm}
   \centerline{ \includegraphics[width=3.5in]{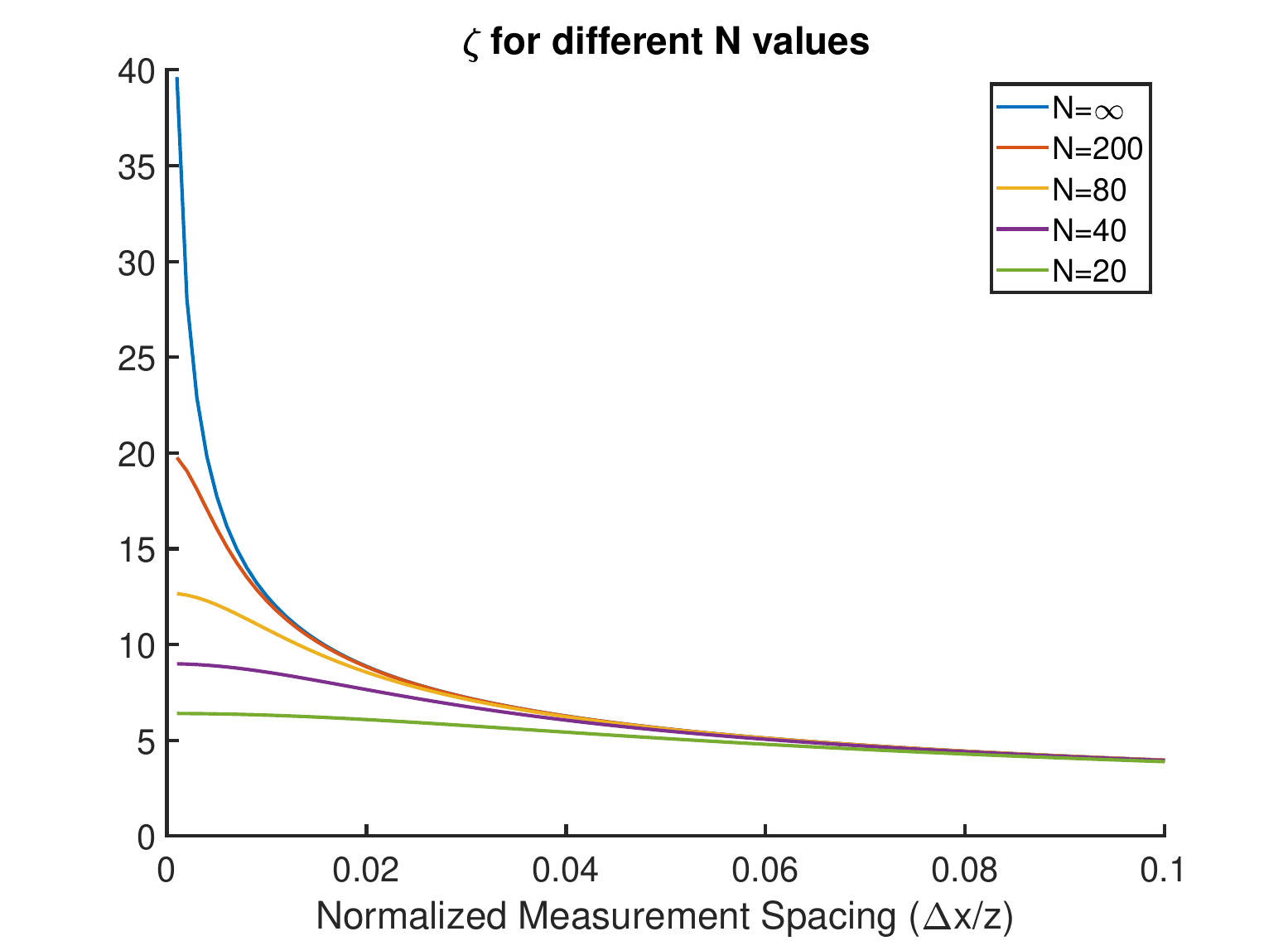}}
\caption{\label{zetafig}
  The top plot shows the value of $\zeta\sub{full}^{\infty}$
  compared to its approximation from~\eqref{zetafullfinal} as a
  function of normalized measurement spacing, $\dd$.   
  The bottom
  plot shows $\zeta\sub{full}$ for different values of $N$ compared to
  the exact limiting case for $N\rightarrow\infty$.  The
  results are very similar to those for $\eta$. 
}
\end{figure}

When only $B_z$ is available the limit of~\eqref{zeta_def} is
\begin{multline}
  \left(\zeta_z^\infty\right)^2=\sum_{i=-\infty}^{\infty}   \left(
     \frac{(1-i^2\dd^2)\cos\theta -
       2i\dd\sin\theta}{(i^2\dd^2+1)^2}\right)^2 =\\
     \frac{3\pi \dd^3\coth(\pi/\dd) +
       3\pi^2\dd^2\csch^2(\pi/\dd)}{12\dd^4}\\
     + \frac{\pi^4\cos2\theta \left( 4\coth^2(\pi/\dd) \csch^2(\pi/\dd)+2\csch^4(\pi/\dd)\right)}{12\dd^4},
\end{multline}
which under the small $\dd$ approximation gives the result
\beq
\zeta_z^{\infty} \approx \sqrt{\frac{\pi}{4\dd}}.
\eeq
This matches the pattern from the other case: $\zeta_z^{\infty}$ is half the value of $\zeta\sub{full}^{\infty}$.
The relative error in the approximation is $+18\%$ when $\dd=1$ and $+4.6\%$ when $\dd=0.75$.  The limit for large $\delta$ is
\beq
\lim_{\delta\rightarrow\infty}\zeta_z^{\infty}=\sqrt{\frac{1+\cos 2 \theta}{2}} = |\cos\theta|,
\eeq
which agrees with the single measurement result from above.

\section{Interpretation of the Bounds}\label{sec:interpretation}
In this section we describe several interpretations of the above results relevant to making design decisions for MCI devices to be employed for tasks of varying complexity. The single point measurement at $x=0$ gives rise to bounds for the case of two
wires with parallel current:
\beq
   s>\sqrt{ \frac{2 \pi \alpha \sigma z^3}{\mu_0 I}}\qquad\text{and}\qquad
   z<\sqrt[3]{\frac{\mu_o I s^2}{2\pi\alpha\sigma}}.
\eeq
For the case of anti-parallel current we have the bounds:
\beq
s > \frac{2 \pi \alpha \sigma z^2}{\mu_0 I} \quad\text{and}\quad
z<\sqrt{\frac{\mu_0Is}{2\pi\alpha\sigma}}.
\eeq
If we repeat this measurement $N$ times with independent noise, then
we can compute the average, whose standard deviation will be 
$\sigma/\sqrt{N}$.  For our problem, the multi-measurement case is analogous, but
the measured quantity is different for each measurement, which
leads to a more complicated form for the effective combined noise
expression.  We can regard the effective noise
being given in the parallel case by
\beq
   \sigma\sub{eff}=\frac{\sigma}{\eta},
\eeq
where $\eta$ is defined in~\eqref{eta_expression},
and in the anti-parallel case by
\beq
   \sigma\sub{eff}=\frac{\sigma}{\zeta},
\eeq
where $\zeta$ is defined in~\eqref{zeta_expression}.

We are interested in the case where the number of magnetic field
measurements, $2N+1$, is large, and when $N\rightarrow\infty$ we have found
\begin{gather}  \label{etanoise}
   \eta\sub{full}^{\infty} \approx \sqrt{\frac{3 \pi}{8\dd}} \quad\text{and}\quad
   \eta_z^{\infty} \approx \sqrt{\frac{3 \pi}{16\dd}} 
   \\  \label{zetanoise}
   \zeta\sub{full}^{\infty} \approx \sqrt{\frac{\pi}{2\dd}}
   \quad\text{and}\quad
   \zeta_z^{\infty} \approx \sqrt{\frac{\pi}{4\dd}},
\end{gather}
where $\delta=\Delta x/z$.  
First note that the $z$-only measurements contain exactly half as many
values as the full measurements, and the corresponding $\eta_z$ and
$\zeta_z$ are $\sqrt{2}$ smaller than those for the full
measurements, which increases the effective noise in~\eref{etanoise}
and~\eref{zetanoise} by $\sqrt{2}$, precisely as expected in general
when we have half as many measurements.  

Decreasing $\dd$ by a factor of two doubles the
measurement density, and hence is equivalent to
doubling the number of measurements.  This change to $\dd$ results in
an increase by a factor of $\sqrt{2}$ of $\eta$ or $\zeta$ and hence
a decrease in the effective noise by that same factor, again
aligning with expectations.

The constant factors in the expressions for $\eta^{\infty}$ and
$\zeta^{\infty}$ are a result of the total effect of multiple
measurements in reducing the effective noise.  These constants depend
on the relative scale of the measurements and the number of
measurements that contribute significantly.  Measurements in the far
field are so close to zero that their contribution is negligible.  As
previously noted, $\eta^{\infty}$ and $\zeta^{\infty}$ both approach
the single measurement value in the limit as $\dd\rightarrow\infty$.  The sensor spacing
required to achieve $\eta\sub{full}^\infty=1.05$ is $\delta\approx 1.3$ and the
spacing to achieve $\zeta\sub{full}^\infty=1.05$ is $\delta\approx 3.0$.  
This means that full field measurements where $|x|>1.3z$ are far enough away to have little effect on detection in the parallel case, and measurements where $|x|>3z$ are expected to have little effect in the anti-parallel case.

Understanding the bounds as a noise adjustment provides a theoretical
interpretation of the bounds, but the form of the bounds is a poor
match to experimental setups.  In particular, the bound has a
hidden dependence on the sensor offset, $z$, through $\dd=\Delta
x/z$.  For a given instrument, the spacing $\Delta x$ is likely to
stay constant even as the instrument standoff varies, but changing the
instrument standoff at constant $\Delta x$ changes $\delta$.  

For the $N\rightarrow\infty$ limit we can reformulate the bounds to
make the results applicable to experiment by
substituting in the approximations for $\eta^{\infty}$ and
$\zeta^{\infty}$, which gives for the parallel case,
\beq
\begin{aligned}
   s&>\sqrt{ \frac{4\sqrt{2\pi} \alpha \sigma}{ \sqrt{3} \mu_oI}}
   (\Delta x)^{1/4} z^{5/4},\\
   z&<\left(\frac{\sqrt{3} \mu_0I}{4\sqrt{2\pi}\alpha\sigma}\right)^{\!2/5}\!\!\!\frac{s^{4/5}}{(\Delta x)^{1/5}},
   \end{aligned}
\eeq
and for the anti-parallel case,
\beq
\begin{aligned}
s &> \frac{2 \sqrt{2\pi} \alpha \sigma}{\mu_0I} (\Delta x)^{1/2} z^{3/2},\\
z&<\left( \frac{\mu_oI}{2\sqrt{2\pi}\alpha\sigma}\right)^{\!2/3}
\!\!\!\frac{s^{2/3}}{(\Delta x)^{1/3}}.
\end{aligned}
\eeq
These bounds lack a clear physical interpretation, but apply directly
to experimental setups.

\begin{figure}
  \centerline{\includegraphics[width=2.8in]{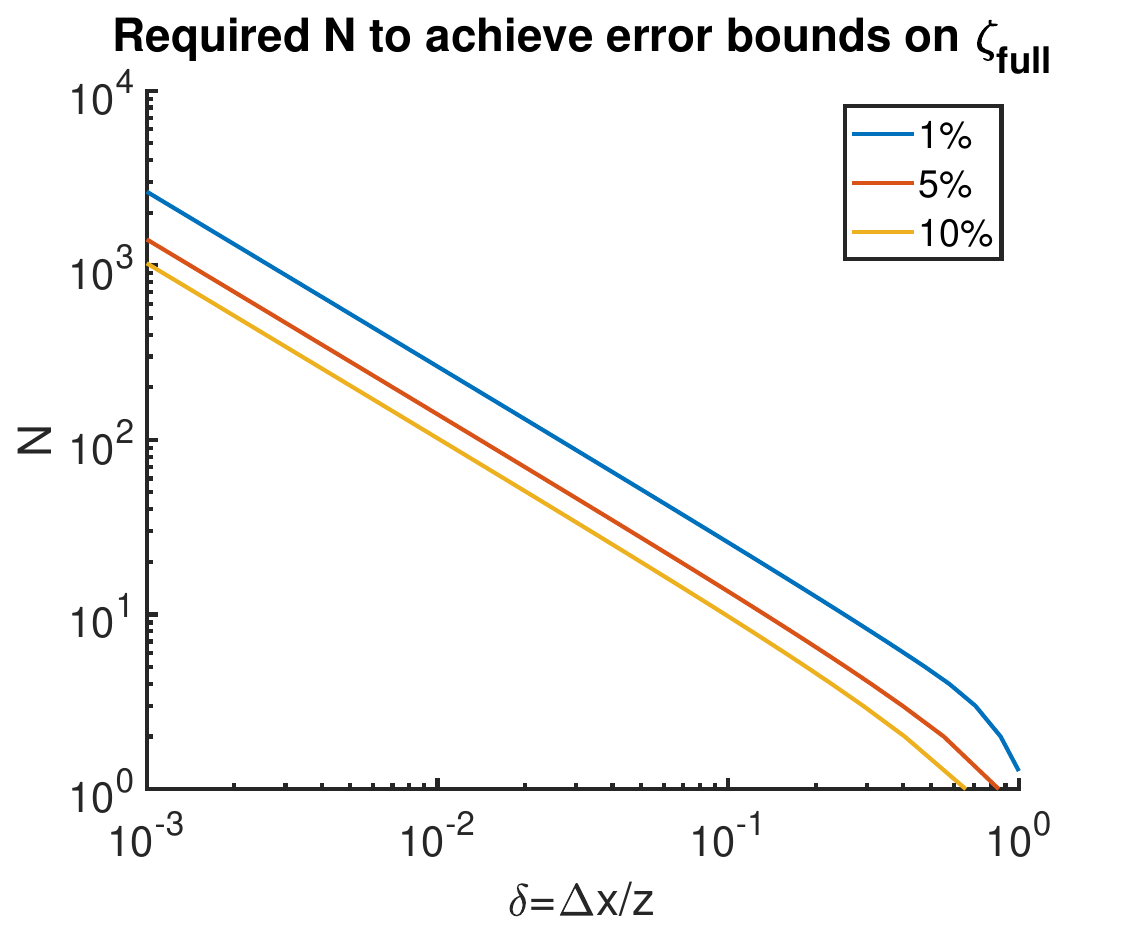}}
  \vspace{0.3cm}
  \centerline{\includegraphics[width=2.8in]{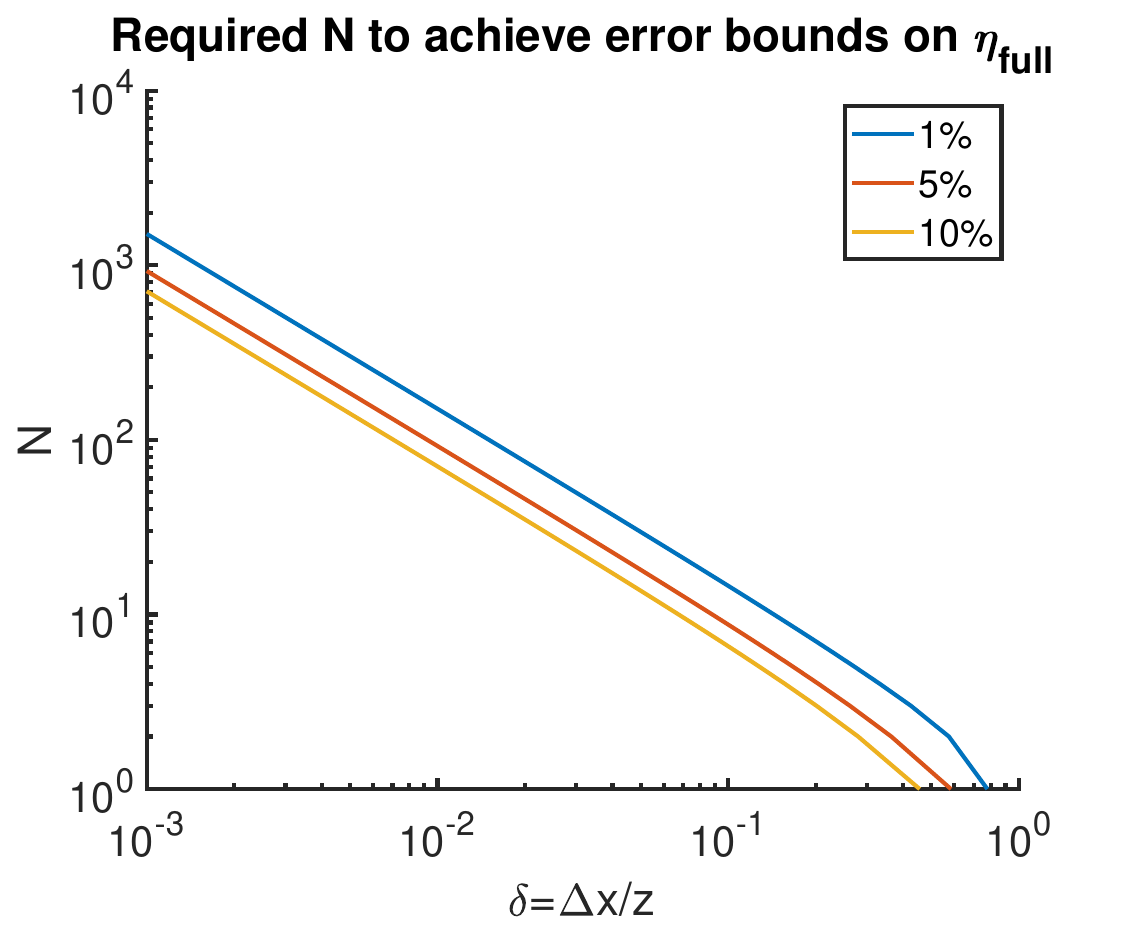}}
  \caption{The required $N$ as a function of $\dd$ for the large $N$ approximation
    to give the specified error when calculating $\eta\sub{full}$ and
    $\zeta\sub{full}$.  These curves are well approximated by
    $C/\delta$ for appropriate $C$.
    \label{Nreq}}
\end{figure}

\begin{figure}
  \centerline{\includegraphics[width=3in]{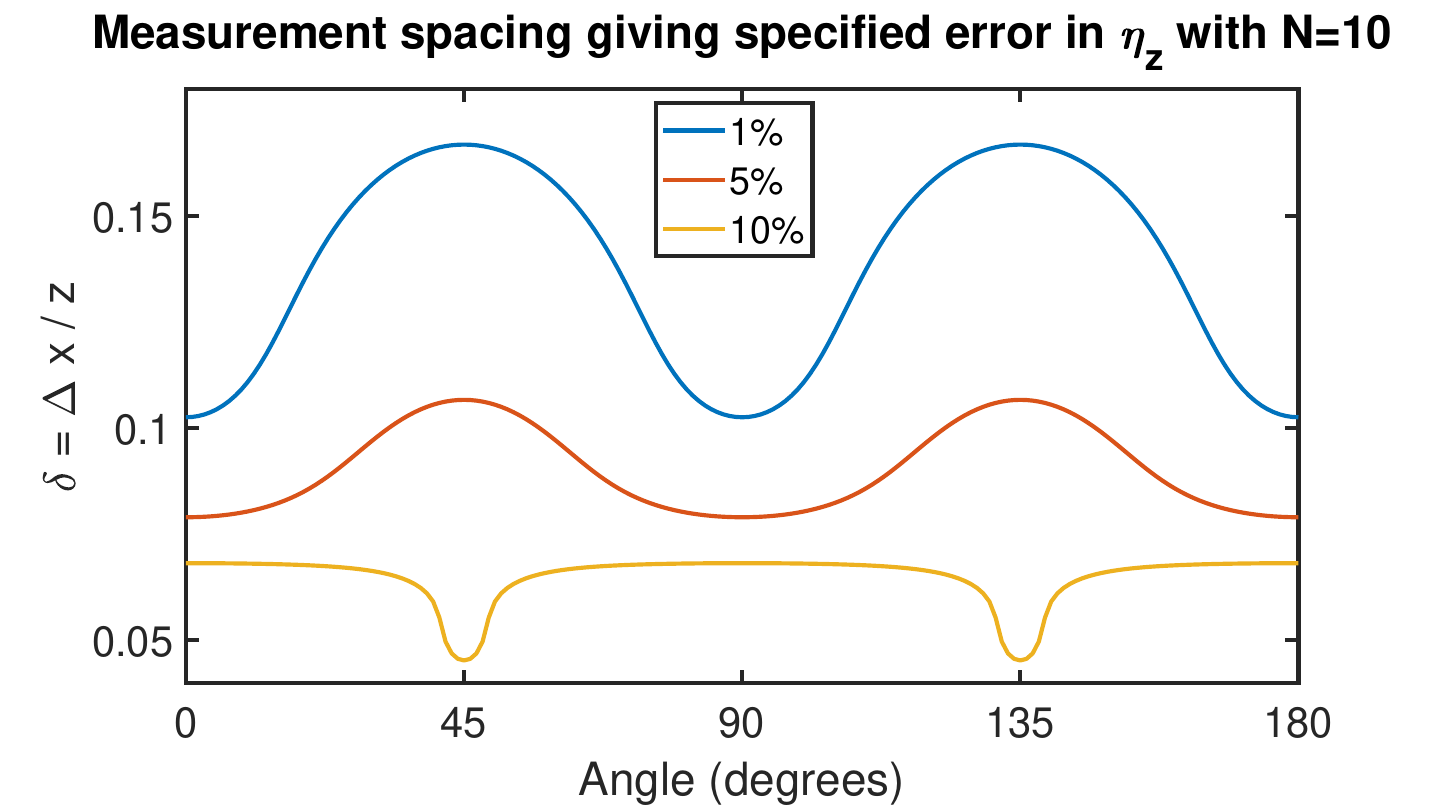}}
  \vspace{0.3cm}
  \centerline{\includegraphics[width=3in]{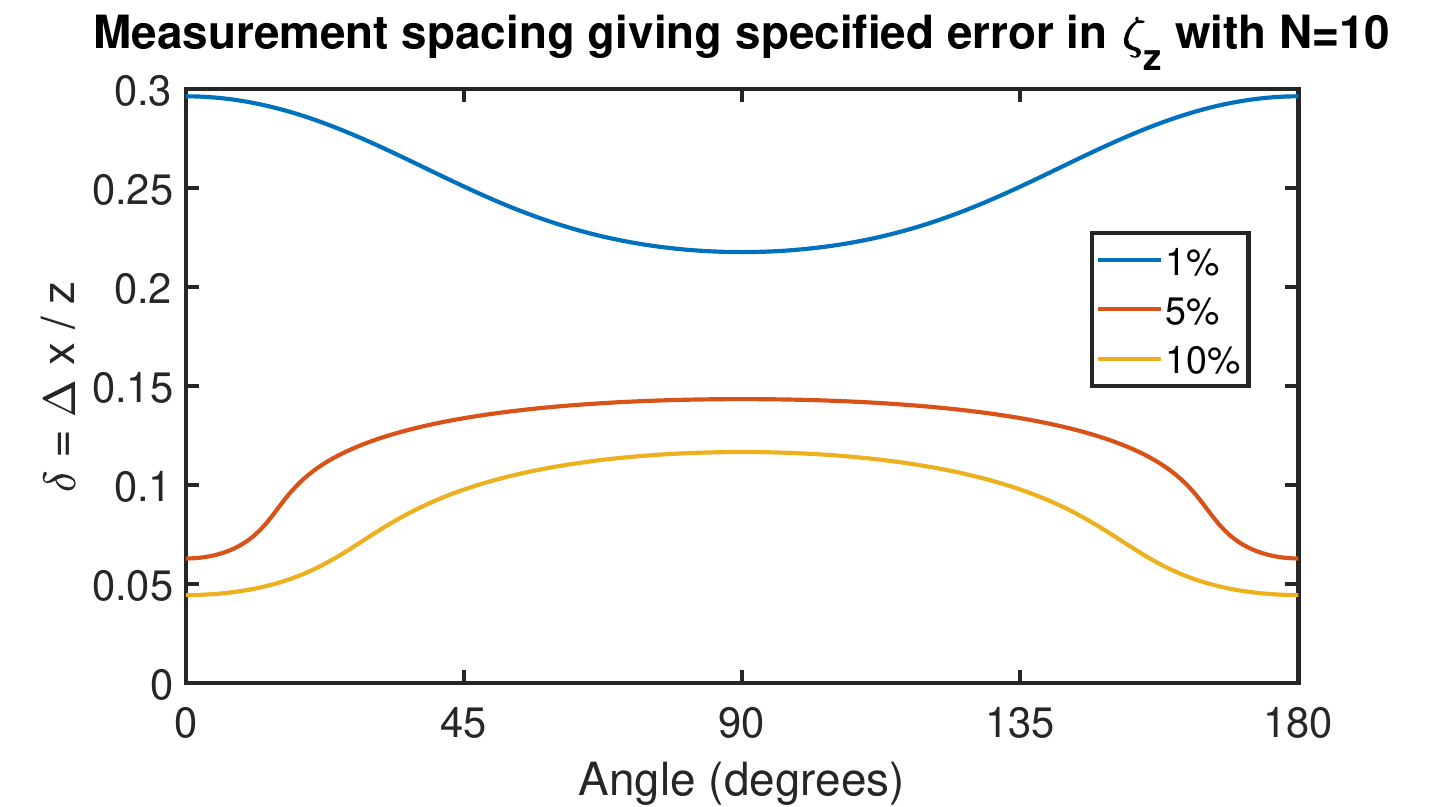}}
  \caption{  \label{thetaworst}
    The plots show the $\dd$ value required in order to achieve a
    specified relative error in $\eta_z$ or $\zeta_z$ as a function of
    angle $\theta$.  The worst case situation occurs when $\dd$ is
    maximal, which happens at $\dd=0^{\circ}$, $\dd=45^{\circ}$, or
    $\dd=90^{\circ}$ depending on the case under consideration. 
  }
\end{figure}

How big does $N$ need to be for the large $N$ approximation to hold?
To answer this question we plot in Figure~\ref{Nreq} the $N$ that is
required as a function of $\dd$ so that 
approximations in~\eref{etanoise} and~\eref{zetanoise} provide a
specified relative error.  
The required $N$ can be bounded below by $C/\dd$, with different values of
$C$ for each case.  When considering the scalar case, where we have
measured only $B_z$, the expressions for $\eta$ and $\zeta$ depend on
$\theta$, so finding a bound on $N$ requires identification of the
worst case $\theta$.  The graphs in Figure~\ref{thetaworst} show that the worst
case $\theta$ value depends on whether we consider the parallel or anti-parallel case, 
but also on the specific relative error we want to achieve.    
Table~\ref{Ctable} gives values of $C$ that
provide a lower bound on $N$ for three error levels, independent of the wire angle.  

\begin{table}
  \caption{\label{Ctable} Choosing $N>C/\dd$ or equivalently
    $N>Cz/\Delta x$ will
    ensure that the the relative error is smaller than the specified
    error value}
  \vspace*{2mm}
  \centering
\begin{tabular}{ccccc}
& \multicolumn{4}{c}{$C$}\\ \cline{2-5}
Error & $\eta\sub{full}$   &   $\zeta\sub{full}$   &   $\eta_z$ & $\zeta_z$\\  \hline
$1\% $&   1.14 &   2.64&  1.76  &    3.12 \\
$5\% $&   0.93 &   1.41&  1.13  &    1.51 \\
$10\%$&   0.71 &   1.03&  0.72  &    1.23 \\ \hline
\end{tabular}
\end{table}

\clearpage
 We finish by providing example bounds on wire separation for some
specific parameter values of interest when measuring with the QDM. We
consider the following values for current flow, which span larger
feature legacy chips to state-of-the-art devices:
\begin{itemize}
  \item  5 mA, the current in a simple integrated circuit with larger
    feature sizes, e.g. the metal layers of a 555
    timer \cite{Kehayias2022}. 
  \item 150 $\mu$A, the current measured in JTAG traces of an 8 nm process
    node NVIDIA GPU with flip chip die \cite{Oliver2021}. 
  \item 100 nA, the current with power delivered to only a few
    active transistors, e.g. wire-bonded 28 nm process node Xilinx
    Artix-7 field programmable gate array with single active ring
    oscillator consisting of six switching transistors \cite{Turner2020}. 
  \item  10 nA, an order of magnitude estimate for the current delivered to a single 
     transistor, based on the previous measurement.  
\end{itemize}
We assume a fixed measurement spacing of $3\micron$ with $N=300$, so
we have 601 measurements across the QDM.  The standoff distance ($z$) can vary
depending on how the chip can be measured.  
\begin{itemize}
\item 800 $\micron$ is typical when the sensor measures an intact
  chip. In this case $\dd=0.00375$.  
\item 200 $\micron$ can be achieved when the chip is thinned, with the epoxy packaging
  milled to decrease packaging thickness, but the components are still
  encapsulated.  In this case $\dd=0.015$.  
\item 1 $\mu$m is possible if the chip is decapsulated so the sensor can rest
  directly on the metallization layer.  This process is difficult and
  the chip is often destroyed. In this case, $\dd=3$.  
\end{itemize}
A final check on $N$ is needed to confirm that the large $N$
approximation holds.  We need $N$ to be the largest when $\dd$ is
small, so we consider the smallest $\dd$ value, $\dd=0.00375$.  In
this case the $5\%$ error bound, $N>248$, holds for vector measurements in
the parallel case. For scalar measurements the $10\%$ error 
bound is $N>275$, which is also satisfied.  The scalar measurements
require a larger $N$, and while we meet the requirement for $10\%$
error in the parallel case, the anti-parallel case requires $N>328$,
which fails.  When we consider the larger $\dd$ values arising at the
closer standoff distances, we find that the $1\%$ error condition is
satisfied in all cases. 

\begin{figure}
\centerline{  \includegraphics[width=3.25in]{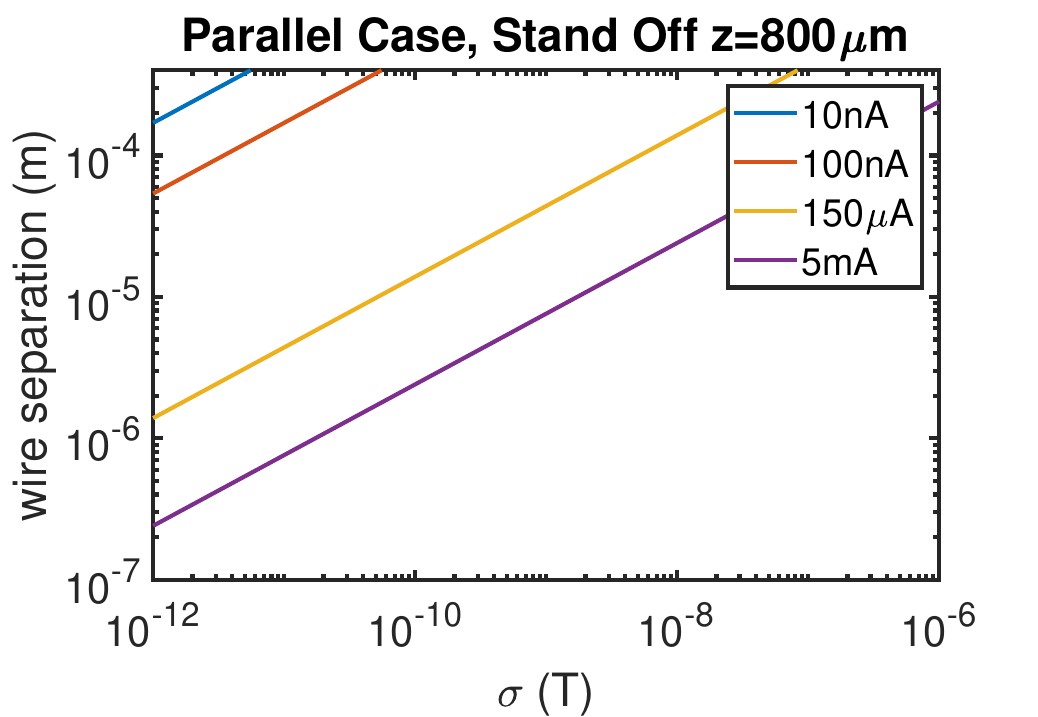}}
\centerline{  \includegraphics[width=3.25in]{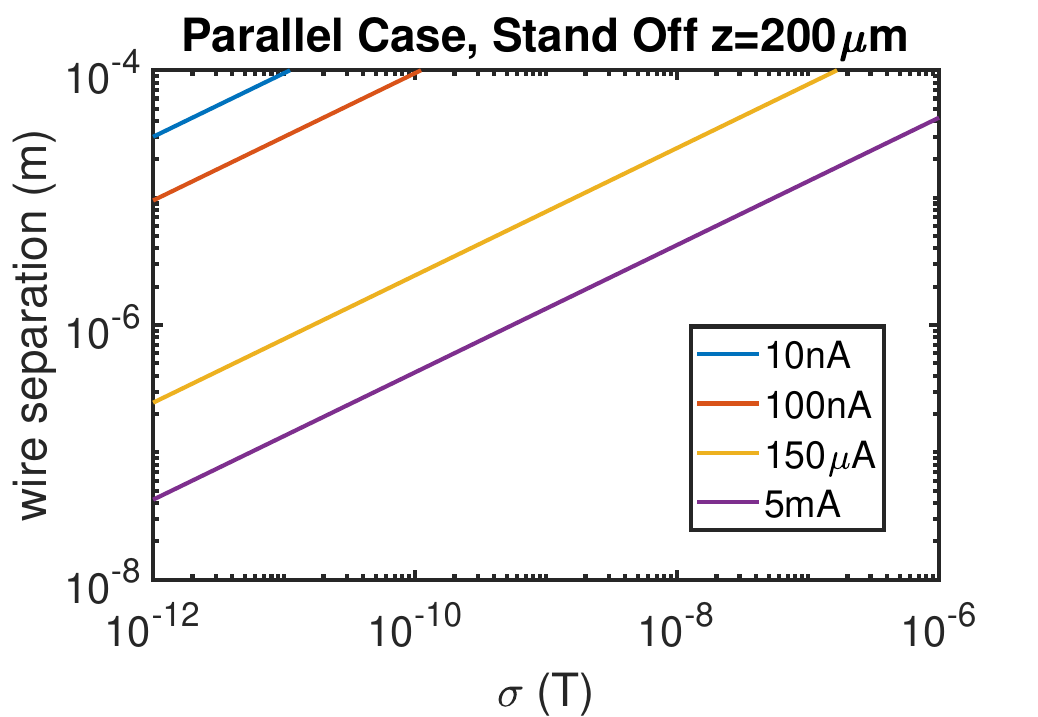}}
\centerline{  \includegraphics[width=3.25in]{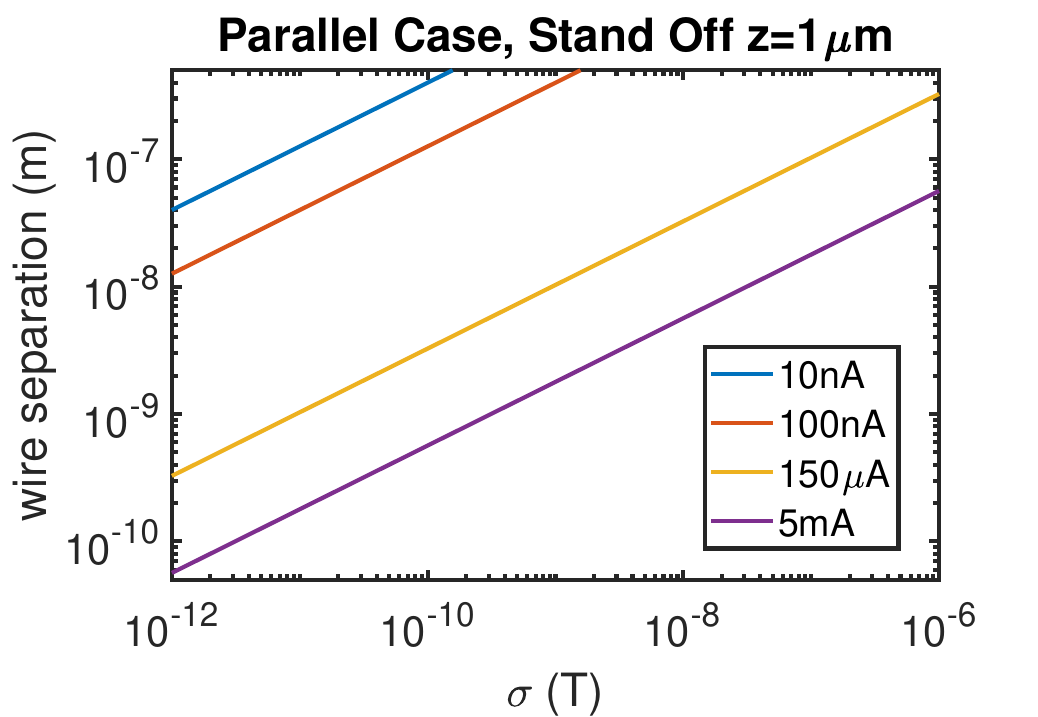}}
  \caption{\label{parbound}
    Plots show the minimum wire separation that can be detected for
    the parallel case, with measurement spacing $3\micron$ and the
    detection threshold parameter, $\alpha=2$.
    The top edge of the
    plots are set to $s=z/2$ so we don't violate the requirement $s\ll z$.
    }
\end{figure}

\begin{figure}
\centerline{\includegraphics[width=3.25in]{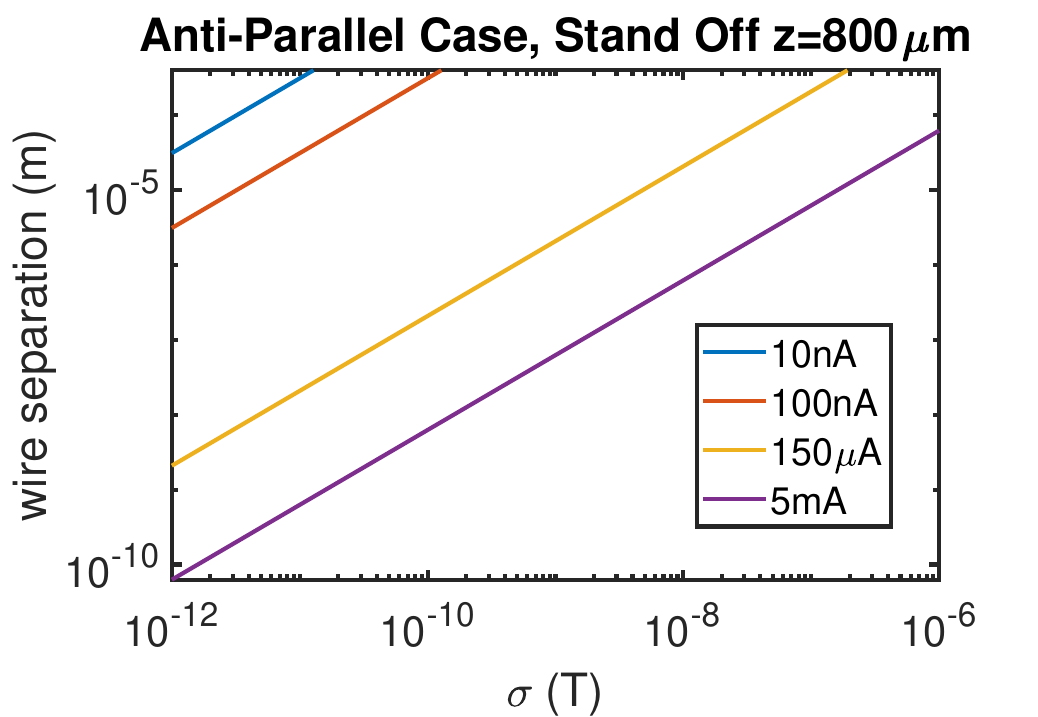}}
\centerline{  \includegraphics[width=3.25in]{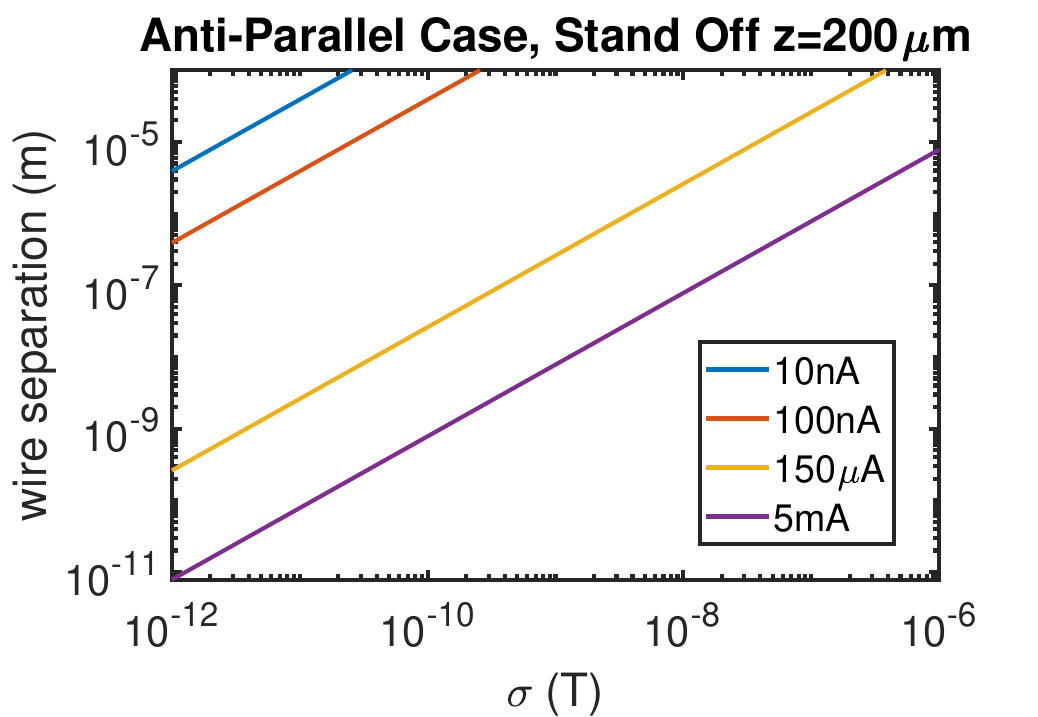}}
\centerline{  \includegraphics[width=3.25in]{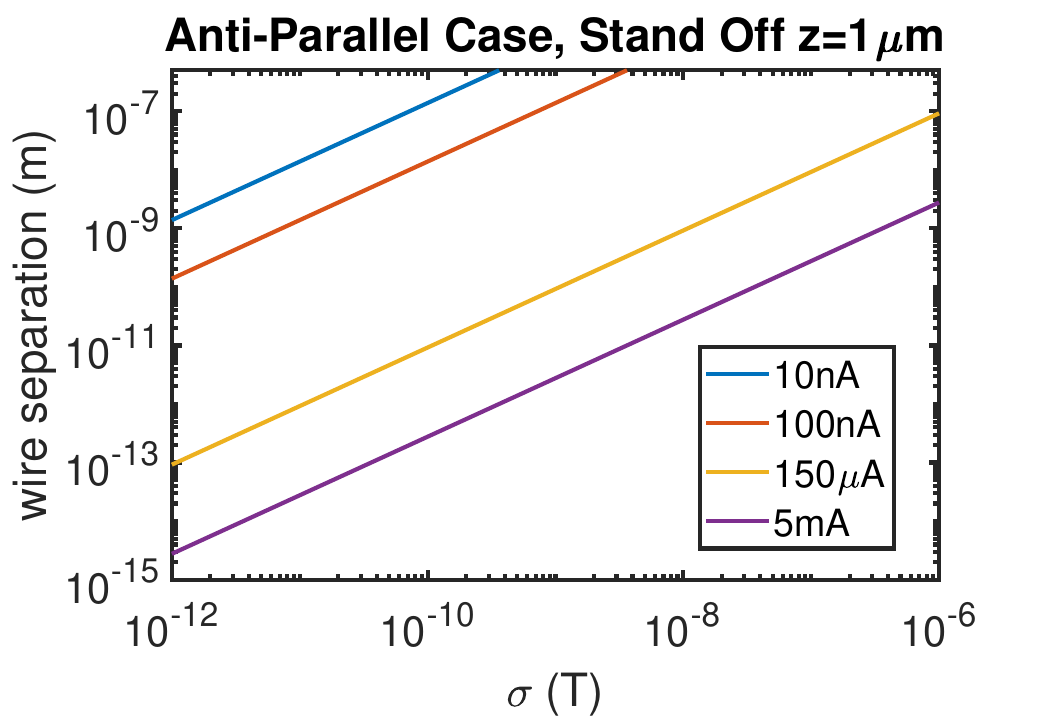}}
  \caption{\label{antibound}
    Plots show the minimum wire separation that can be detected for
    the anti-parallel case, with measurement spacing $3\micron$ and the
    detection threshold parameter, $\alpha=2$.  The top edge of the
    plots are set to $s=z/2$ so we don't violate the requirement $s\ll z$.
    }
\end{figure}

Plots of bounds on the wire spacing ($s$) for measurement of the full
vector magnetic field appear in
Figure~\ref{parbound} for the parallel case and Figure~\ref{antibound}
for the anti-parallel case.
The bound on $s$ can be understood as a bound on spatial resolution, and
is based on the assumption that
$s\ll z$;  we show data up to $s=z/2$ to stay within this constraint.
In the scalar case, where only $B_z$ is available, we can use the same
plots but simply adjust $\sigma$ to $\sqrt{s}\sigma$.

\section{Conclusions}\label{sec:conclusions}
In order to understand the utility of Magnetic Current Imaging (MCI) when applied to microelectronics inspection tasks, we have presented analysis of the simplest possible magnetic current imaging tasks: distinguishing a pair of wires from a single wire and distinguishing a pair of wires from empty space.  The analysis provides lower bounds on the separability of two current carrying wires as a function of current, measurement noise, wire separation, and measurement standoff distance. As this represents the simplest possible MCI task, this analysis provides a useful bound on the utility of MCI applied to specific imaging tasks. These bounds can be employed to estimate the feasibility of more challenging MCI tasks. We applied these formula to determine wire separability via MCI as a function of measurement noise for current levels and measurement standoff distances that are representative of real devices and existing measurement protocols.

\section{Acknowledgements}
The view, opinions, and/or findings in this report are those of The MITRE Corporation and should not be construed as an official Government position, policy, or decision, unless designated by other documentation. This technical data was developed using contract funds under Basic Contract No. W56KGU-22-F-0017.
Approved for Public Release; Distribution Unlimited. Public Release Case Number 24-0461.
\textcopyright~2024 The MITRE Corporation. 

\bibliography{bounds}

\begin{IEEEbiographynophoto}{Adrian Mariano}
received the B.S. degree in mathematics from the University of Washington in 1993 and the Ph.D. degree in applied mathematics from Cornell University in 1999.  Since then he has worked at the MITRE Corporation, investigating topics connected to hyperspectral imagery and electromagnetic sensing.   
\end{IEEEbiographynophoto}

\begin{IEEEbiographynophoto}{Jacob Lenz}
 received the B.S. degree in physics from University of Wisconsin--Milwaukee in 2020 and the M.S. degree in Physics--Quantum Computing from the University of Wisconsin--Madison in 2021, where he aided in the setup of a lab exercise using NV centers in diamond. Since then he has worked as a quantum physicist at the MITRE Corporation, carrying out experimentation with the Quantum Diamond Microscope, which utilizes NV centers in diamond to detect activity in microelectronics components.  
\end{IEEEbiographynophoto}

\begin{IEEEbiographynophoto}{Dmitro Martynowych}
 received the B.S. degree from The University of Scranton in 2015 in Chemistry and the Ph.D. degree from The Massachusetts Institute of Technology in 2020 in Physical Chemistry with a thesis focused on the generation and spectroscopic probing of shockwaves and other high pressure-temperature phenomena. From 2021 to 2023, he was a Physicist with the MITRE Corporation where he developed techniques for magnetic field sensing with defect centers in diamond.  He currently holds an appointment as a research scientist in the Laboratory for Information and Decision Systems (LIDS) at MIT.
\end{IEEEbiographynophoto}

\begin{IEEEbiographynophoto}{Christopher Miller}
 received the B.S. degree in mathematics from Fairfield University in 2006 and the Ph.D. degree in applied mathematics in 2012. He specializes in numerical analysis, high performance computing, computational imaging, and remote sensing.  
\end{IEEEbiographynophoto}

\begin{IEEEbiographynophoto}{Sean Oliver}
 received both the B.S. and Ph.D. degrees in physics from George Mason University in 2014 and 2020, respectively, with a specialization in materials science. Following graduation, he joined the MITRE Corporation where his expertise lies in quantum sensing, specifically in the area of magnetic imaging using NV centers in diamond. 
\end{IEEEbiographynophoto}

\end{document}